\documentstyle[prb,aps]{revtex}
\title{Plastic energies in layered superconductors}
\author{N.K. Wilkin \cite{ic}}
\address{Department of Physics,
University of Sheffield,\\ Sheffield, S3 7RH, United Kingdom.}
\author{M.A. Moore}
\address{Department of Theoretical Physics, University of Manchester,\\
Manchester, M13 9PL, United Kingdom}
\date{\today}
\draft
\begin{document}
\maketitle

\begin{abstract}

We estimate the energy cost associated with two pancake vortices
colliding in a layered superconductor. It is argued that this energy
sets the plastics energy scale and is the analogue of the crossing
energy for vortices in the continuum case. The starting point of the
calculation is the Lawrence-Doniach version of the Ginzburg-Landau
free energy for type-II superconductors. The magnetic fields
considered are along the c-direction and assumed to be sufficiently
high that the lowest Landau level approximation is valid. For Bi-2212,
where it is know that layering is very important, the results are
radically different from what would have been obtained using a
three-dimensional anisotropic continuum model.  We then use the
plastic energy for Bi-2212 to successfully explain recent results from
Hellerqvist {\em et al.}\ on its longitudinal resistance.

\end{abstract}

There has been much recent experimental interest in the dependence of
the longitudinal resistivity in the HTC-superconductors
\cite{Hellerqvist,Wan94,Safar94} on field and temperature. In a recent
paper, we produced a theory for some aspects of this data which
involved knowing the crossing energy for two vortices (flux
lines). Our work was essentially an extension of the hydrodynamic
theory of Marchetti, Nelson and Cates
\cite{Marchetti,Nelson93,Cates}. It gave a reasonable fit to the
longitudinal resistivity data of Safar {\em et al.}\ \cite{Safar94} on YBCO
crystals.  However, it was clear from our calculation that the length
scale over which the crossing took place was of the order of the
spacing between the superconducting Cu-O planes. Our calculation had
been done using anisotropic continuum Ginzburg-Landau theory, and so
neglected any layering effects. The crossing energy that we calculated
for the three-dimensional anisotropic case can be thought of as the
plastic energy cost of making non-trivial topological deformations of
the vortices, and we shall refer to it as $U_{\rm pl}^{\rm cont}$.

In this paper we calculate the energy associated with two pancakes
colliding within a layer: this is an estimate of the plastic
deformation energy for vortices in a strongly layered superconductor
and sets the timescales for deformation of the vortices within the
system. Intuitively one might have expected that neighbouring pancake
vortices in a layer would have exchanged positions by moving around
each other. However, within the Lowest Landau Level (LLL)
approximation which we use, Dodgson and Moore \cite{Dodgson94} have
shown that the lowest energy barrier for this to occur is when the
vortices collide and pass through each other.

 We start from a Lawrence-Doniach Ginzburg-Landau free energy
functional We find that including the layering leaves the results for
the plastic energy of YBCO essentially unchanged but gives markedly
different results for Bi-2212. This suggests that treating YBCO as a
3D continuum anisotropic superconductor within GL theory is
reasonable, but that in Bi-2212 the effect of the Cu-O layers must be
included --- as was to be expected as the coupling between the layers
in Bi-2212 is known to be very weak.

To fix notations, we briefly describe the modified Ginzburg-Landau
theory for a layered superconductor, where $\psi$, is our spatially
dependent order parameter.  We begin with the continuum free energy functional,
\begin{equation}
\label{vo:e:lg1}
\frac{F[\psi(r)]}{k_{B}T_{c}}=\int \mbox{d}^{3}r \left(\alpha(T)\,
|\psi|^2+\beta\,\frac{|\psi|^{4}}{2} + \sum_{\mu=1}^3
\frac{|(-i\hbar\partial_{\mu}-2e\mbox{\bf A}_{\mu})\psi|^{2}}{2
m_{\mu}}+\frac{B^2}{2\mu_0}\right)\: .
\end{equation}
Here $\alpha (T)$ is the temperature-dependent variable, $\beta$ is
the coupling constant, and $m_{\mu}$ are the effective masses. In the
cases we consider the masses in the $ab$-plane are taken as equal and
are denoted by $m_{ab}$, and the mass in the c-direction is written as
$m_c$. The temperature dependence of $\alpha(T)$ is taken to be of the
form , $\alpha(T)= (T-T_{c}) \alpha'$. We also assume the LLL approximation
(see Refs.~\onlinecite{RT,Wilkin} for
justification of it). In this approximation the derivative terms in
the direction perpendicular to the planes reduce to $e \hbar B/m_{ab}
|\psi|^2$, so it is convenient to write the temperature variable as
$\alpha_{H} =\alpha+e B\hbar/m_{ab}$. This is zero along the
$H_{c_{2}}$ line, negative below and positive above. According to
Te\u{s}anovi\'{c} et al.\ \cite{Tesanovic} the LLL approximation can
be trusted when $H > H_{c_2}/f$, where $f=3$.

In our previous calculation we chose to consider a system that
contained only two vortices, but with different types of periodic
boundary conditions. In this paper we instead generalise to the
layered case the procedure of Dodgson and Moore \cite{Dodgson94} who
allowed only the two vortices that are crossing to move and kept all
the other vortices fixed in their triangular lattice positions. The
advantage of using a background of fixed vortices is that the
resulting crossing energy, $U_{\rm pl}$ is an upper-bound on the true
crossing energy. Furthermore, it has been shown that the corrections
to this bound seem to be very small, at least for the continuum limit.
\cite{Dodgson94}.

The general form for the order parameter in the $ab$-plane within the LLL
approximation is:
\begin{equation}
\psi(x,y)=f(\zeta)e^{-\pi |\zeta|^2 B/(2 \Phi_0)} \qquad {\rm where}\qquad
\zeta=x+ i y,
\end{equation}
and $f(\zeta)$ can be written in the product form,
\begin{equation}
\label{cr:gl:genfor}
f(\zeta) \propto \Pi_i (\zeta -\zeta_i)
\end{equation}
where the ${\zeta_i}$ are the (complex) positions of the vortices. The
minimum potential free energy configuration for this system is the
triangular lattice of lattice spacing $\sqrt{3}/2 l^2=\Phi_0/B$. We
will label the order-parameter that describes this configuration by
$\psi_0(x,y)$.

We are interested in two vortices being displaced along the $x$-axis
from their lattice positions at $\zeta=\pm l/2$ and colliding at the
origin $x=y=0$. Using the formalism of Ref.~\onlinecite{Dodgson94} we
have for the order parameter at height $z$
\begin{equation}
\label{phiM}
\psi(x,y,z)=\psi_0(x,y)\frac{(\zeta+a(z))(\zeta-a(z))}{(\zeta^2-l^2/4)}
\end{equation}
$2 a(z)$ is the
separation of the colliding vortices and $a(0)=0$.
(see Fig.~\ref{fig:coord}).

Layering will be taken into account by replacing in Eq.~\ref{vo:e:lg1}
the derivative $\partial \psi(x,y,z) /\partial z$ by
$(\psi_{n+1}(x,y)-\psi_n(x,y))/d$ where $\psi_n(x,y)$ denotes the
value of $\psi$ in the $n^{\rm th}$ layer at point $(x,y)$ and $d$ is
the effective spacing between the superconducting Cu-O layers. In the
context of the LLL formalism, such an approximation has been derived
as a consequence of a periodically varying $T_c(z)$ by \v{S}\'{a}\v{s}ik and
Stroud \cite{Sasik93}. Note that within the LLL approximation the
vector potential ${\bf A}$ is always that of the uniform field so
${A_z} \equiv 0$

The resulting effective Ginzburg-Landau free energy functional within the LLL
approximation for a layered structure can then be written as,
\begin{equation}
\label{GLlayer}
F=d\, \sum_n \int\left( - |\alpha_H||\psi_n|^2
+\frac{\beta}{2} |\psi_n|^4+
\frac{\hbar^2}{(2 m_c d^2)}|\psi_{n+1}-\psi_n|^2\right) {\rm d}x \,{\rm d} y
\end{equation}

The previous results for the continuum case of Dodgson and Moore
\cite{Dodgson94}, with a rigid lattice background of vortices led to a
crossing energy of $2.3 \,\hbar \Phi_0\, |\alpha_H|^{3/2}/(\sqrt{2
m_c}\,\beta B)$,which is a rigorous upper bound to the crossing
energy. Our earlier calculations using only two vortices with periodic
boundary conditions led to an energy of $1.46\,\hbar \Phi_0 \,
|\alpha_H|^{3/2}/(\sqrt{2 m_c}\, \beta B)$, which is somewhat lower,
possibly due to errors from the finite size of the system. In this
paper we use the rigid lattice approach, as the equations are more
readily physically interpreted, and easier to solve.

The extra free energy, $\Delta F$ associated with the two displaced
vortices as compared with the original lattice is found by
substituting Eq.~(\ref{phiM}) into Eq.~(\ref{GLlayer}) and then doing
the integral over the $(x,y)$ directions \cite{Dodgson94},

\begin{equation}
\label{eq:fdisc}
\Delta F= \frac{2 d|\alpha_H|^2 \Phi_0}{\sqrt{3} \beta \beta_A
B}\,g(\eta) \; \; \; \mbox{where} \; \; \;
g(\eta)=\sum_{n=-N}^{N}\left
(\overbrace{\left(\sum_{m=0}^{4}c_m a_n^{2m}\right)}^\alpha +
\overbrace{\frac{I_1}{\eta^2} (a_{n+1}^2 -a_n^2)^2}^\beta\right)
\end{equation}

where $a_n$ is the value of $a(z)$ in the n$^{\rm th}$ layer of which
there are $(2N+1)$. The factor $\beta_A\simeq 1.16$ is the Abrikosov
factor. The quantity $\eta=d/\xi_c$ where $\xi_c=\hbar/\sqrt{2
m_c|\alpha_H|}$ --- the `$c$-axis correlation length' measures the
importance of layering effects.  The formula for $\xi_c$ in terms of
$\alpha_H$ should be modified \cite{RT} very close to the $H_{c_2}$
line a regime which we therefore shall not consider. When $\eta$ is
large, layering effects are important; $\eta
\rightarrow 0$ is the continuum limit. The constants $c_m$ and $I_1$
are determined by integration in the $xy$-plane, $c_0=0.7903$,
$c_2=-5.201$, $c_3=-32.7365$, $c_4=100.8232$, and $I_1=23.5502$.  The
$\alpha$ and $\beta$ terms can be respectively thought of as the
`potential' and `kinetic' contributions to the plastic energy.

In order for the vortices to collide we require that at the central
layer ($n=0$) the displaced vortices lie on top of each
other. By symmetry this should occur at $a_0=0$. (See
Fig.~\ref{fig:coord}.) Our other boundary condition is that $a_{\pm
\infty}=\pm l/2$, that is we require the vortices to return to their
equilibrium positions in the original lattice.

The energy barrier for vortices to collide is then found by minimising
Eq.~(\ref{eq:fdisc}), with the boundary conditions imposed. The
results unlike in the continuum case cannot be written in terms of a
reduced temperature variable. They are now field, temperature and
layer thickness $d$ dependent. The easiest way to get an idea of how
the plastic energy in the layered case varies is to compare it with
the plastic/crossing energy calculated for the rigid lattice background in the
continuum case.

\begin{equation}
\label{eq:comp}
\frac{U_{\rm pl}^{\rm disc}}{U_{\rm pl}^{\rm cont}}= 0.43 \eta g(\eta)
\end{equation}

where $U_{\rm pl}^{\rm disc}$ is the crossing energy found in the
discrete case and $U_{\rm pl}^{\rm cont}$ in the continuum
case. The form of this function is shown in
Fig.~\ref{fig:comp}. Putting in typical system parameters for YBCO:
$|\mbox{d} B/\mbox{d} T|=2$T/K, $m_{ab}/m_c=\epsilon^2=1/59$,
$T_c=93$K, and a spacing between Cu-O planes of 11.4$\AA$ (where we
have treated the closely spaced pair of Cu-O layers as a `single'
layer) we find that $\eta \sim 2$, for values of field and temperature
in the vicinity of the irreversibility line. As can be seen from
Fig.~\ref{fig:comp} for $\eta \sim 2$ the plastic energy calculated
according to the continuum approximation and that calculated allowing
for layering are approximately the same, indicating that for YBCO the
continuum approximation is satisfactory, certainly for all the regions
of the phase diagram in which we would expect the LLL approximation to
be valid. The material for which layering effects would be expected to
be more important is Bi-2212, where it is known from experiments that
the coupling between the Cu-O layers is very weak.  Our calculation
bears this out; using system parameters of $|{\rm d} B/{\rm d}T|=0.5$T/K,
$T_c=86$,
$\epsilon=1/55$ (there appears to be a factor of $1/3$ uncertainty in the value
of $\epsilon$!) and $d=30.9
\AA$, we find $\eta \sim 25$ and that the
$U_{\rm pl}^{\rm disc}/U_{\rm pl}^{\rm cont}\sim 8$. Thus the continuum
estimate appears to seriously
underestimate the plastic energy for this weakly coupled layered
system.

Analysis of $g(\eta)$ shows that for the values of $\eta$ appropriate
to Bi-2212 we can greatly simplify the formula for the plastic energy,
Eq.~(\ref{eq:fdisc}). The displacement of the pancake vortices for
these values of $\eta$, is completed within three layers (see
Fig.~\ref{fig:cross}), and is such that the `kinetic energy' term in
Eq.~(\ref{eq:fdisc}) is negligible in comparison to the `potential
energy' term, which tends to the constant $c_0$.  Hence when $\eta$ is
large
\begin{equation}
U_{\rm pl}^{\rm disc}\simeq \frac{2 d |\alpha_H|^2 \Phi_0}{\sqrt{3} \beta
\beta_A B} \, c_0.
\end{equation}

In the three dimensional continuum case we were able to use the formula of
Cates
{\em et al.}\ \cite{Marchetti,Nelson93,Cates} which says that the length scale
over which the system is coherent along the field direction,
$\delta_c$ can be written as
\begin{equation}
\label{eq:deltac}
\delta_c \sim l_e \exp[U_{\rm pl}^{\rm cont}/k T].
\end{equation}
The distance $l_e$ is the distance along a flux line that is traveled
before encountering another flux line, and is related to the
equilibrium flux line spacing by $l_e=a_0^2/l_p$, where $l_p$ is a
persistence length \cite{Cates} which in our case is equal to the
c-axis correlation length $\xi_c$. (There is uncertainty in the
numerical constant multiplying $l_e$ but it should be of O(1)). We
motivated Eq.~(\ref{eq:deltac}) in Ref.~\onlinecite{wilkin3} in some
detail but it can also be readily understood using an explanation of
Marchetti and Nelson \cite{Marchetti91}. If we concentrate on the
dynamics of one flux line in a flux liquid and label the c-direction
by time then $1/l_e$ becomes an effective attempt frequency to cross or
collide,
which occurs with probability $\exp[-U_{\rm pl}^{\rm cont}/kT]$, leading
immediately to
Eq.~(\ref{eq:deltac}). Crossing of flux lines always produces a loss
of phase coherence. In the layered case we can no longer consider
`crossing energies'. However, the arguments detailed above follow
through once we consider the plastic deformation energy rather than
the crossing energy. The importance of the plastic energy for flux line motion
in the liquid state is shown in Blatter {\em et al.}\ \cite{Blatter}.

We are now in a position to compare with the longitudinal resistivity
measurements of Hellerqvist {\em et al.}\ \cite{Hellerqvist} on Bi-2212. Using
a single crystal of size $2\times 2\times 0.030$mm (smallest dimension
in the c-direction) they measured the dynamic resistance $\partial
V/\partial I$ as a function of bias current in a magnetic field along
the c-direction. Their data of $\rho_c$ against $T$ has a smooth
peaked behavior with the position of the peak shifting to higher
temperatures as the field is lowered. By fitting to a power law of
the form $\partial V/\partial I \propto I^\alpha$ they find linear
$I-V$ behavior for $T>T_{\rm peak}$ but for $T<T_{\rm peak}$ the
exponent $\alpha$ is positive and varies smoothly with temperature and field.

The apparent disappearance of the longitudinal linear resistivity at
the peak suggest that
the sample has become phase coherent along the
c-direction.
To locate the point at which the longitudinal resistivity vanishes
we set $\delta_c$ to equal the system size, $L=0.03$mm.

Analysis of the Hellerqvist {\em et al.}\ \cite{Hellerqvist} data, shows that
the field-temperature dependence of the longitudinal resistivity peaks
is well fitted by the formula $L\simeq \delta_c$ (see
Fig.~\ref{fig:exptal}). At high fields and low temperatures we begin
to see deviations from the experimental data --- but this is no
surprise, as  these points are so far away from the $H_{c_2}$ line
that it is unlikely that the LLL approximation remains valid.

We conclude by observing that the onset of vanishing longitudinal
resistance is well-explained by equating the plastic energy length scale
$\delta_c$ to the system size. No phase transition
\cite{Feigelman} need be invoked to explain the experimental data.

We should like to thank the Institute for Theoretical Physics at Santa
Barbara for for their hospitality during the writing of this paper
(NSF grant PHY89-04035). We are grateful to Matthew Dodgson for the use of his
numerical results, Aharon Kapitulnik for telling us of his
experimental data and Monica Hellerqvist for the experimental
data. One of us (N.K.W) was supported by the SERC through grant number
GR/H9427.

\newpage
\begin{figure}
\caption{Trajectory of two lines colliding in the $xy$-plane}
\label{fig:coord}
\end{figure}

\begin{figure}
\caption{The ratio $U_{\rm pl}^{\rm disc}/U_{\rm pl}^{\rm cont}$ versus
$\eta=d/\xi_c$. As $\eta \rightarrow 0$ the ratio tends to unity while
it is linear for large $\eta$.}
\label{fig:comp}
\end{figure}

\begin{figure}
\caption{Configuration for vortices colliding in Bi-2212.
The positions of the vortices are only defined in the layers}
\label{fig:cross}
\end{figure}

\begin{figure}
\caption{Comparison of experiment (points) and theory (solid) line for the
 line marking the disappearance of the longitudinal resistivity in Bi-2212.}
\label{fig:exptal}
\end{figure}


\begin{thebibliography}{10}

\bibitem{ic}
Currently at Department of Mathematics, Imperial College, London, SW7 2BZ, UK.

\bibitem{Hellerqvist}
M.~C. Hellerqvist, S. Ryu, L.~W. Lombardo, and A. Kapitulnik, Physica C {\bf
  230},  170  (1994).

\bibitem{Wan94}
Y.~M. Wan, S.~E. Hebboul, and J.~C. Garland, Phys.\ Rev.\ Lett. {\bf 72},  3867
   (1994).

\bibitem{Safar94}
H. Safar, P.~L. Gammel, D. Huse, S.~N. Majumdar, L.~F. Scneemeyer, D.~J.
  Bishop, D. Lop\'{e}z, G. Nieva, and F. de~la Cruz, Phys.\ Rev.\ Lett. {\bf
  72},  1272  (1994).

\bibitem{Marchetti}
M.~C. Marchetti and D.~R. Nelson, Phys.\ Rev.\ B {\bf 42},  9938  (1990).

\bibitem{Nelson93}
D.~R. Nelson,  in {\em Phase Transitions and Relaxation in Systems with
  Competing Energy Scales}, Vol.~415 of {\em NATO ASI Series C}, edited by T.
  Riste and D. Sherrington (Kluwer Academic Publishers,
  Dordrecht/Boston/London, 1993).

\bibitem{Cates}
M.~E. Cates, Phys.\ Rev.\ B {\bf 45},  12415  (1992).

\bibitem{Dodgson94}
M.~J.~W. Dodgson and M.~A. Moore, phys.\ Rev.\ B in press.

\bibitem{RT}
G.~J. Ruggeri and D.~J. Thouless, J.\ Phys.\ F {\bf 6},  2063  (1976).

\bibitem{Wilkin}
N.~K. Wilkin and M.~A. Moore, Phys.\ Rev.\ B {\bf 47},  957  (1993).

\bibitem{Tesanovic}
Z. Te\u{s}anovi\'{c}, L. Xing, L.~B.~Q. Li, and M. Suenaga, Phys.\ Rev.\ Lett.
  {\bf 69},  3563  (1992).

\bibitem{Sasik93}
R. {\v{S}\'{a}\v{s}ik} and D. Stroud, Phys.\ Rev.\ B {\bf 48},  9938  (1993).

\bibitem{wilkin3}
M.~A. Moore and N.~K. Wilkin, Phys.\ Rev. B {\bf 50},  10  (1994).

\bibitem{Marchetti91}
M.~C. Marchetti and D.~R. Nelson, Physica C {\bf 174},  40  (1991).

\bibitem{Blatter}
G. Blatter, M.~V. Feigel'man, V.~B. Geshkenbein, A.~I. Larkin, and V.~M.
  Vinoku, Rev.\ Mod.\ Phys. {\bf 66},  1125  (1994).

\bibitem{Feigelman}
M.~V. Feigel'man, V.~B. Geshkenbein, L.~B. Ioffe, and A.~I. Larkin, Phys.\ Rev.
  B {\bf 48},  16641  (1993).

\end{thebibliography}
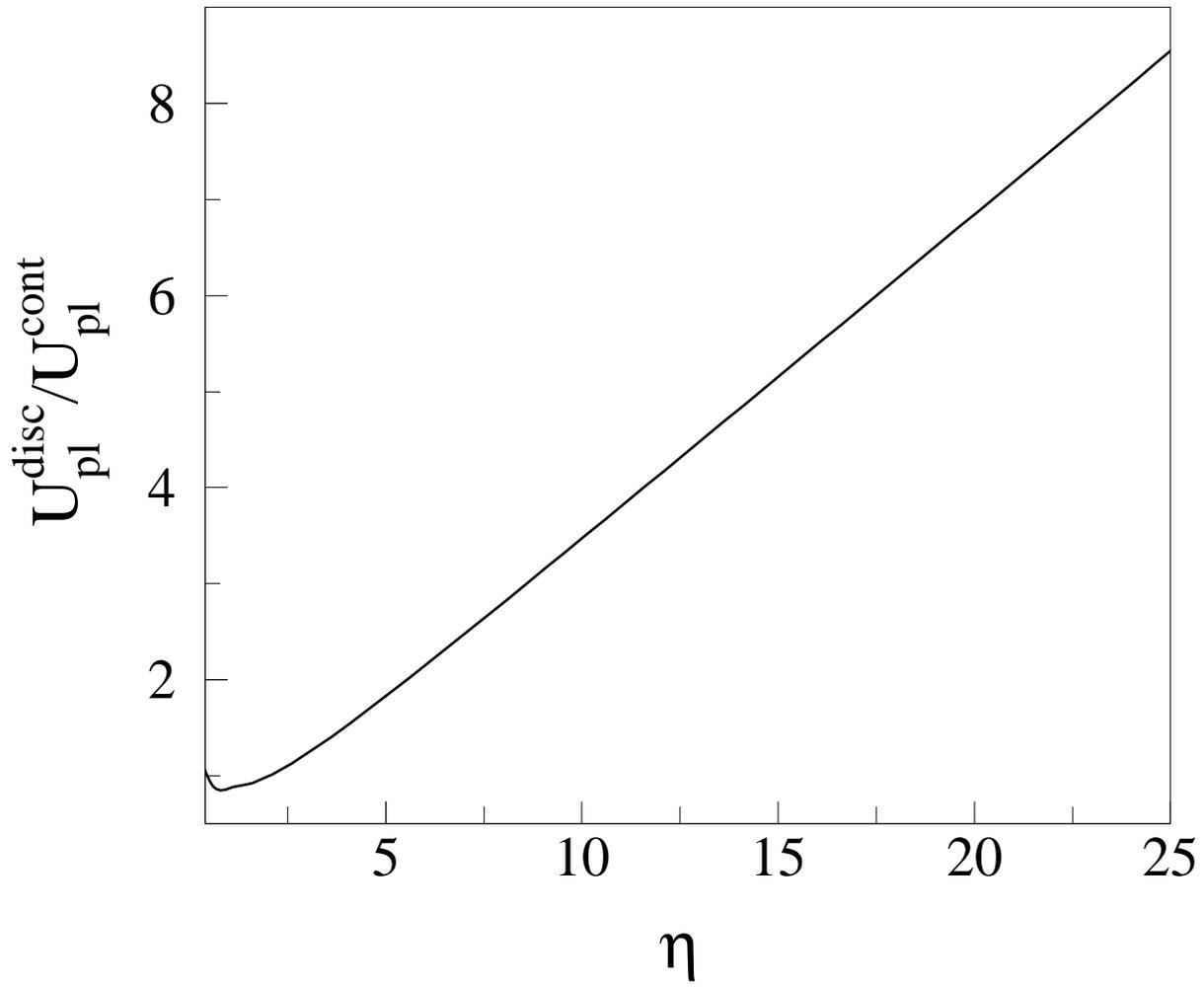
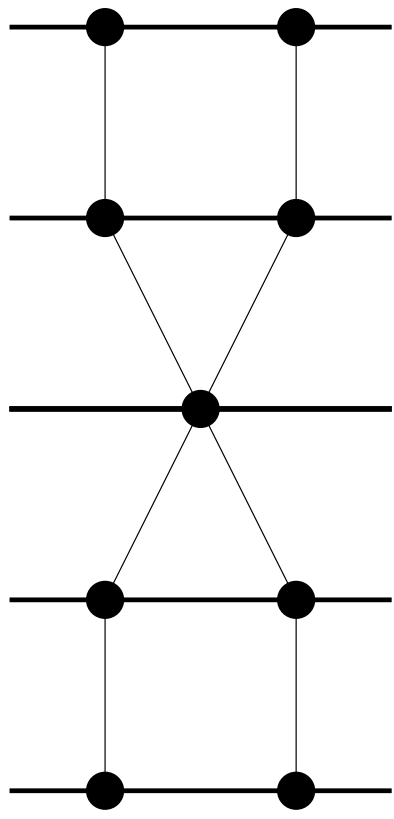
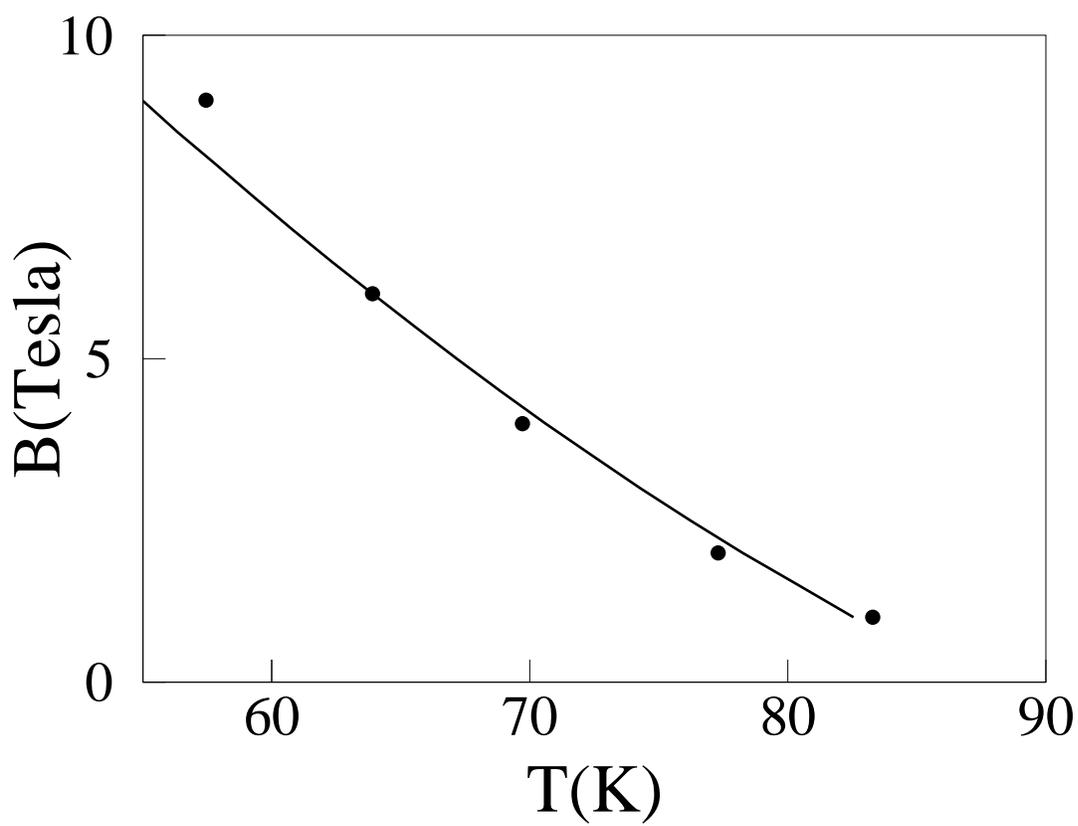
\end{document}